\documentclass[11pt,twoside]{article}
\usepackage{newpasp}

\usepackage{epsf}

\index{Heckman, T. M.}

\begin{document}

\title{Galactic Superwinds at Low and High Redshift}
\author{Timothy M. Heckman}
\affil{Department of Physics \& Astronomy, Johns Hopkins University,
Baltimore, MD 21218}

% A concise abstract is recommended.  Enter the text of the abstract in
% between the \begin{abstract} and \end{abstract} commands.  Do NOT
% include the word ``Abstract'' in your text; it is inserted
% automatically. Do NOT  make a paragraph break between \begin{abstract} 
% and the first line of the text of the abstract!  Abstracts are required 
% for all papers.

\begin{abstract}
In this contribution I summarize our current knowledge of the nature
and significance of starburst-driven galactic superwinds. These flows
are driven primarily by the kinetic energy supplied by supernovae.
Superwinds are complex, multiphase phenomena requiring a panchromatic
observational approach. They are ubiquitous
in galaxies in which the global star-formation rate per unit area
exceeds roughly 10$^{-1}$ M$_{\odot}$ yr$^{-1}$ kpc$^{-2}$
(a condition satisfied by local starbursts and high-z Lyman Break
galaxies). Data on X-ray emission, optical line-emission, and optical/UV
interstellar absorption-lines together imply that the mass outflow rates
are comparable to the star-formation-rates and that the conversion
of kinetic energy from supernovae to superwind is quite efficient
($\sim$ 30 to 100\%). Measured/inferred outflow speeds range from
a few $\times 10^2$ to 10$^3$ km/s and appear to be independent of the rotation speed
of the ``host'' galaxy. The outflows are dusty (dust/gas ratios
of $\sim$ 1\% by mass). These properties imply that 
superwinds may have established the mass-metallicity relation
in elliptical and bulges, polluted the inter-galactic
medium to a metallicity of $\sim$ 10 to 30\% solar, heated the
inter-galactic medium by up to $\sim$1 kev per baryon, and ejected
enough dust into the inter-galactic medium to have potentially observable 
consequences. 

\end{abstract}

% Include keywords if you wish. The keywords.apj file, found on aas.org 
% in the pubs/aastex-misc directory, contains a list of keywords used 
% with the ApJ and Letters.  

%%\keywords{infrared: galaxies -- galaxies: nuclei -- galaxies: starburst}

% That's it for the front matter.  On to the main body of the paper.

\section{Introduction}
By now, it is well-established that galactic-scale outflows of gas (sometimes
called `superwinds') are commonplace in the most actively star-
forming galaxies in both the local universe (e.g. Heckman et al 2000;
Dahlem, Weaver, \& Heckman 1998; Lehnert \& Heckman 1996)
and at high redshift (Franx et al 1997; Pettini et al 1998).
They are powered by the energy deposited in the
interstellar medium by massive stars via supernovae and stellar winds.
Over the history of the universe, outflows like these may
have polluted the intergalactic medium with metals (e.g. Nath \& Trentham
1997) and dust (Aguirre 1999), heated and
polluted the intracluster medium
(e.g.
Ponman, Cannon, \& Navarro 1999),
and may have established the mass-metallicity relation and radial metallicity
gradients in galactic spheroids  (e.g. Carollo \& Danziger 1994).
However, the cosmological relevance of superwinds can not be
reliably assessed without first understanding their physical, dynamical, and
chemical properties.

In this contribution, I will summarize the basic theoretical ideas concerning
the energetics and dynamical evolution of superwinds and
describe the likely physical origin of the emission and absorption
they produce (section 2). I will then review the observed properties of
superwinds: their demographics (section 3, their estimated
outflow rates (section 4), and their likely fate (section 5).
Finally, I will briefly describe the potential
implications of superwinds for the evolution of galaxies and the inter-galactic
medium (section 6).

\section{The Conceptual Framework}
The engine that drives the observed outflows in
starbursts is the mechanical energy supplied by massive stars in
the form of supernovae and stellar winds (cf. Leitherer \& Heckman
1995). For typical starburst parameters, the rate
of supply of mechanical energy is of-order 1\% of the bolometric
luminosity of the starburst and typically 10 to 20\% of the Lyman
continuum luminosity. Some fraction of this mechanical energy may
be radiated away by dense shock-heated material inside the
starburst. However, observations of superwinds
imply that a significant fraction is available to drive the outflow
(see below). Radiation pressure acting on dust grains may also play
a role in driving the observed outflows (Aguirre 1999).

The dynamical evolution of a starburst-driven outflow has been
extensively discussed (e.g. Chevalier \& Clegg 1985; Wang 1995;
Strickland \& Stevens 2000).
Briefly, the deposition of mechanical energy by
supernovae and stellar winds results in an over-pressured cavity
of hot gas inside the starburst. The temperature of this hot gas
is given by:
\begin{displaymath}
T = 0.4 \mu m_H \dot{E}/k\dot{M} \sim 10^8 l^{-1} K
\end{displaymath}
for a mass (kinetic energy) deposition rate of $\dot{M}$ ($\dot{E}$).
The ``mass-loading'' term $l$ represents the ratio of the total mass
of gas that is heated to the mass that is directly
ejected by supernovae and stellar winds (e.g. $l \geq$ 1).

This hot cavity will expand, sweep
up ambient material and thus develop a bubble-like structure. If
the ambient medium is stratified (like a disk), the `superbubble' will
expand most rapidly in the direction of the vertical pressure
gradient. After the superbubble size reaches several disk vertical scale
heights, the expansion will accelerate, and it is believed that
Raleigh-Taylor instabilities will then lead to the fragmentation
of the bubble's outer wall. This allows the hot gas to `blow out'
of the disk and into the galactic halo in the form of a weakly
collimated bipolar outflow (i.e. the flow makes a transition from
a superbubble to a superwind). The terminal velocity of this hot
wind is expected to be in the range of one-to-a-few thousand
km s$^{-1}$:
\begin{displaymath}
v_{wind} = (2 \dot{E}/\dot{M})^{1/2} \sim 3000 l^{-1/2} km/s
\end{displaymath}
The observational manifestations of superbubbles and superwinds are
many and varied. The ambient gas (both the material in the outer
superbubble wall and overtaken clouds inside the superbubble or
superwind) can be photoionized by the starburst and shock-heated
by the outflow. This material can produce soft X-rays and
optical/ultraviolet emission and absorption lines.
The predicted expansion speed of the outer wall of an adiabatic wind-blown
superbubble is of-order 10$^2$ km s$^{-1}$:
\begin{displaymath}
v_{Bubble} \sim 100 \dot{E}_{42}^{1/5} n_0^{-1/5} t_7^{-2/5} km/s
\end{displaymath}
for a bubble driven into an ambient medium with nucleon density
$n_0$ (cm$^{-3}$) by
mechanical energy deposited at a rate $\dot{E}_{42}$ (units of
10$^{42}$ erg s$^{-1}$) for a time $t_7$ (units of 10$^7$ years).
Clouds exposed to the ram pressure of the wind will be accelerated
to terminal velocities of few hundred km s$^{-1}$:
\begin{displaymath}
v_{cloud} \sim 600 \dot{p_{34}}^{1/2}\Omega_w^{-1/2}r_{0,kpc}^{-1/2}
N_{cloud,21}^{-1/2} km/s
\end{displaymath}
for a cloud with a column density $N_{cloud,21}$ (units of 10$^{21}$
cm$^{-2}$) that - starting at an initial radius of $r_0$ (kpc) - is
accelerated by a wind that carries a total momentum flux of
$\dot{p_{34}}$ (units of 10$^{34}$ dynes) into a solid angle 
$\Omega_w$ (steradian).

The hot gas that
drives the expansion of the superbubble/superwind may itself be an
detectable source of X-rays, especially if a significant amount of
mass-loading of the outflow occurs in or around the starburst
(e.g. $l >>$ 1). Finally, cosmic ray electrons and
magnetic field may be advected out of the starburst by the flow and
produce a radio synchrotron halo and possibly an X-ray halo via
inverse Compton scattering of soft photons from the starburst (
Seaquist \& Odegard 1991; Moran, Lehnert, \& Helfand 
1999). From both a theoretical and observational
perspective, it is clear that {\it superwinds are a complex, multiphase
phenomenon requiring a panchromatic observational approach.}

\section{Superwind Demographics}
Lehnert \& Heckman (1996) discussed the analysis of the optical emission-line
properties of a sample of $\sim$50 disk galaxies selected to be bright and warm
in the far-infrared (active star-formers) and to be viewed within
$\sim30^{\circ}$ of edge-on
(to facilitate detection of outflows along the galaxy minor axis).
They defined several indicators of minor-axis outflows: 1) an excess of
ionized gas along the minor axis (from H$\alpha$ images) 2) emission-line
profiles that were broader along the galaxy minor axis than along the major
axis 3) emission-line ratios that were more ``shock-like'' along
the galaxy minor axis than the major axis (e.g. had
stronger [OI]$\lambda$6300, [NII]$\lambda$6584, and [SII]$\lambda$
$\lambda$6717,6731 emission relative to H$\alpha$). They found that
all these indicators became stronger in the galaxies with more
intense star-formation (larger $L_{FIR}$, larger $L_{FIR}/L_{OPT}$,
and warmer dust temperatures).

{\it In summary, the optical emission-line evidence implies
that superwinds are ubiquitous in galaxies with star-formation-rates
per unit area $\Sigma_* \geq$ 10$^{-1}$ M$_{\odot}$ yr$^{-1}$ kpc$^{-2}$.
Starbursts surpass
this threshold, while the disks of ordinary spirals do not (Kennicutt 1998).}

Dahlem, Weaver, \& Heckman (1998) used $ROSAT$ and $ASCA$ to search
for X-ray evidence for outflows from a complete sample of the 
seven nearest edge-on starburst galaxies
(selected on the basis of far-IR flux, warm far-IR colors, 
edge-on orientation, and low
Galactic foreground HI column). Apart from the dwarf galaxy NGC55, all the
galaxies showed hot gas in their halos. The gas had temperatures of
a few times 10$^6$ to 10$^7$ K, and could be traced out to distances
of-order 10 kpc from the disk plane. The overall bulk properties
of the X-ray halos (size, energy content, X-ray luminosity) were
consistent with simple models of superwinds
(see also Read, Ponman, \& Strckland
1997). 

Finally, we (Heckman et al 2000 - hereafter H2000) have completed a survey of
the NaI$\lambda$5893 (``NaD'')
absorption-line doublet
in a sample of 32 far-infrared-selected starburst galaxies.
In 18 cases, the line was produced primarily by interstellar gas,
and in 12 of these its centroid was blueshifted by over 100 km/s relative to
the galaxy systemic velocity. The outflows occurred in galaxies
systematically viewed more nearly face-on
than the others.
The absorption-line profiles in these outflow
sources spanned the range from near the galaxy
systemic velocity to a typical maximum blueshift of 400 to 600 km s$^{-1}$,
which we argued represented the terminal velocity reached by ambient
interstellar clouds accelerated along
the minor axis of the galaxy by the hot superwind fluid.

At high-redshift, the only readily available tracers of superwinds
are the interstellar absorption-lines in the rest-frame ultraviolet.
Six of the seven star-forming galaxies at $z =$ 2.7 to 4.9
in the combined samples of
Franx et al (1997), Pettini et al (1998), and
Pettini et al (2000) showed
interstellar absorption-lines that were blueshifted by a few hundred
to over a thousand km s$^{-1}$ relative to the estimated galaxy systemic
velocity. Moreover, a ``composite'' spectrum formed from the sum of the spectra
of 12 star-forming galaxies at $z \sim$ 3
also showed interstellar absorption-lines blueshifted by several
hundred km s$^{-1}$ 
(Franx et al 1997). The high-redshift
galaxies with outflows all easily exceed the threshold in $\Sigma_*$
given above for local galaxies driving outflows (see Meurer et al 1997).

Very similar kinematics are observed in UV spectra
of local starbursts (Heckman \& Leitherer 1997;
Kunth et al 1998; Gonzalez-Delgado et al 1998). $FUSE$ far-UV spectra
recently obtained by our group shows that the outflows contain 
coronal-phase gas ($T \sim$ few $\times$ 10$^5$ K), as probed with
the OVI$\lambda$$\lambda$1032,1038 doublet (Heckman et al and
Martin et al, in preparation).

\section{Estimates of Outflow Rates}

\subsection{X-ray Emission}
While it is relatively straightforward to demonstrate qualitatively that
a superwind is present, it is difficult to reliably calculate the
rates at which mass, metals, and energy are being transported out by
the wind. Several different types of data can be used, each with its
own limitations and required set of assumptions. 

X-ray imaging spectroscopy yields the superwind's
``emission integral'' (the integral over the emitting volume of
the square of the gas density). Presuming that the X-ray spectra
are fit with the correct model for the hot gas it follows that
the mass and energy of the X-ray gas scale as follows:
$M_X \propto (L_X f)^{1/2}$ and 
$E_X \propto (L_X f)^{1/2} T_X (1+{\cal M}^2)$. Here
$f$ is the volume-filling-factor of the X-ray gas and
$\cal M$ is its Mach number. Numerical hydrodynamical simulations 
of superwinds suggest that
$\cal M$$^2$ = 2 to 3 (Strickland \& Stevens 2000).
The associated outflow rates ($\dot{M}_X$ and $\dot{E}_X$)
can then be estimated by dividing
$M_X$ and $E_X$ by the crossing time of the observed region:
$t \sim (R/c_s\cal M)$, where $c_s$ is the speed-of-sound.

If the X-ray-emitting gas is assumed to be volume-filling ($f \sim$
unity), the resulting values for $\dot{E}_X$ are then very similar
to the rate of kinetic energy deposition by the starburst. This
would imply that very little of this energy is lost due to radiative cooling.
Similarly, the estimated values for $\dot{M}_X$ significantly
exceed the rate at which massive stars directly return mass to the ISM:
that is, the outflow has to be strongly ``mass-loaded'' with ambient
interstellar gas ($l \sim$ 5 to 10, typically). The implied outflow
rates are then similar to the estimated star-formation rates
(e.g. Martin 1999).

The high spatial resolution data provided by $Chandra$
is proving to be quite instructive in testing the assumptions described
above. In particular, observations of the outflow in the prototypical
starburst/superwind galaxy NGC253 (Strickland et al 2000) show that
the X-ray emission in the outflow is limb-brightened and
coincident with the H$\alpha$ filaments (Figure 1). We argue that the X-rays
arise at the turbulent interface (mixing layer) between a very fast, hot, and
tenuous wind fluid
(whose X-ray emission is undetectably faint) and the walls of the hollow
cavity carved by this wind. On morphological and physical grounds we
argue that $f$ is of-order 10$^{-1}$ for the {\it observed} X-ray-emitting
gas. If this is true in general, it would mean that previous estimates
of $\dot{M}_X$ and $\dot{E}_X$ are overestimated by a factor of $\sim$ 3.
The majority of the outflow's energy would reside in
a largely invisible fluid (shades of dark matter!).

\begin{figure}
\plotone{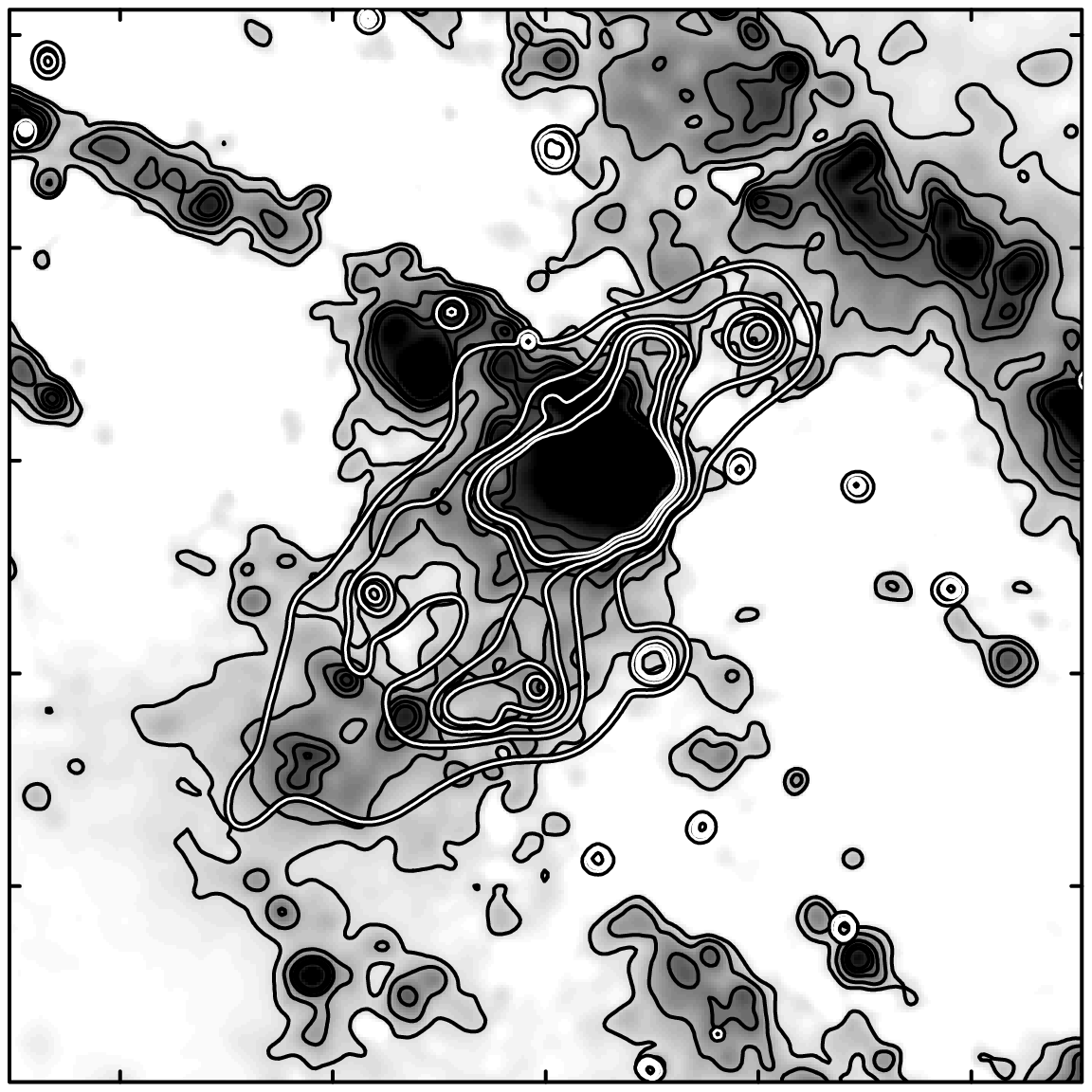}
\caption{\small $Chandra$ ACIS soft (0.3 to 2 keV) image of the center
of NGC253 shown in white contours and a logarithmically-scaled H$\alpha$
image in grey-scale. Tic marks are separated by 30 arcsec (380 pc). Note
the similarity between the two images, and the strong limb-brightening
in the southeast ``outflow cone''. See Strickland et al (2000).}
\label{fig1}
\end{figure}

In order to estimate a rate at which metals are transported out,
we have to know the metallicity of the hot gas. It seems clear
that the bulk of the mass of the X-ray emitting material is provided
by ambient material that has been heated in some way by the superwind.
Thus, we would expect this material to have roughly solar abundances.
This is at odds with some X-ray analyses (e.g. Ptak et al 1997),
but my colleagues and I would argue that the current X-ray
data are indeed consistent with rather normal $\sim$solar abundances
(Weaver, Heckman, \& Dahlem 2000; Strickland \&
Stevens 2000). The superior capabilities of the $Chandra$ and
$XMM-Newton$ X-ray observatories may be able to settle this matter.

\subsection{Optical Emission}

Optical data on the warm ($T \sim$ 10$^4$ K) ionized gas can be used
to determine the outflow rates $\dot{M}$ and $\dot{E}$ in a way that
is quite analogous to the X-ray data. 
In this case, the outflow velocities
can be directly measured kinematically from spectroscopy. They range
from $\sim10^2$ km s$^{-1}$ in starbursting dwarf galaxies 
(Marlowe et al 1995; Martin 1998) to a few $\times$ 10$^2$ to 10$^3$
km s$^{-1}$ in powerful starbursts (e.g. Heckman, Armus, \& Miley
1990). Martin (1999) found the implied values for $\dot{M}$
are comparable to (and may even exceed) the star-formation rate.

In favorable cases, the densities and thermal pressures can be directly
measured in the optical emission-line clouds using density- and
temperature-sensitive ratios of emission lines. The thermal pressure in these
clouds traces the ram-pressure in the faster outflowing wind that is
accelerating them (hydrodynamical simulations suggest that
$P_{ram} = \Psi P_{cloud}$, where $\Psi$ = 1 to 10). Thus, for a wind with a
mass-flux $\dot{M}$ that freely flows
at a velocity $v$ into a solid angle $\Omega$, we have
\begin{displaymath}
\dot{M} = \Psi P_{cloud}\Omega r^2/v
\end{displaymath}
\begin{displaymath}
\dot{E} = 0.5 \Psi P_{cloud}\Omega r^2 v
\end{displaymath}
Based on observations and numerical models, the values
$v \sim$ 10$^3$ km s$^{-1}$, 
$\Psi \sim$ a few, and $\Omega/4\pi \sim$ a few tenths are reasonable.
The radial pressure profiles $P_{cloud}(r)$ measured in superwinds by
Heckman, Armus, \& Miley (1990) and Lehnert \& Heckman (1996) then imply
that $\dot{M}$ is comparable to the star-formation rate (requiring
$l$ = 5 to 10) and that $\dot{E}$ is comparable to the starburst
kinetic-energy injection rate (implying that radiative losses are
not severe).

Rigorously determined metallicities for the optical emission-line
material are not available. However, comparisons of the observed
spectra to shock or photoionization models imply that the abundances
are consistent with those in the ambient ISM of the host galaxy
(e.g. subsolar in dwarfs and roughly solar in the more massive galaxies).

\subsection{Interstellar Absorption-Lines}

The use of interstellar absorption-lines to determine outflows rates
offer several distinct advantages.
First, since
the gas is seen in absorption against the background starlight, there is no
possible ambiguity as to the sign (inwards or outwards) of any radial flow
that is detected, and the outflow speed can be measured directly.
Second, the strength of the absorption will be related  to the
column density of the gas. In contrast, the X-ray or optical surface-brightness
of the emitting gas is proportional to the emission-measure. Thus, the
absorption-lines more fully probe the whole range of gas densities in the
outflow, rather than being strongly weighted in favor of the densest material
(which may contain relatively little mass).  Finally, provided that suitably-
bright background sources can be found, interstellar absorption-lines have been
used to study outflows in high-redshift galaxies where the associated X-ray
or optical {\it emission} may be undetectably faint (see above).

The biggest obstacle to estimating outflows rates is that the strong
absorption-lines are usually saturated, so that their equivalent width
is determined by the velocity dispersion and covering factor, rather
than by the ionic column density. In the cases where the rest-UV region
can be probed with adequate signal-to-noise (Pettini et al 2000; Heckman
\& Leitherer 1997),
the total $HI$ column in the outflow can be measured
by fitting the damping wings of the Lyman~$\alpha$ interstellar line,
while ionic columns may be estimated from the weaker (less saturated)
interstellar lines. In the H2000 survey of the
$NaD$ line, we estimated $NaI$ columns in the outflows based
on the $NaD$ doublet ratio (Spitzer 1968), and we then
estimated the $HI$ column assuming that the gas obeyed the same
relation between $N_{HI}$ and $N_{NaI}$ as in the Milky Way. These
$HI$ columns agreed with columns estimated independently from the line-of-sight
color excess $E(B-V)$ toward the starburst, assuming a Galactic gas-to-dust
ratio. From both the UV data and the $NaD$ data, the typical inferred
values for $N_{HI}$ are of-order 10$^{21}$ cm$^{-2}$.

We can then adopt
a simple model of a constant-velocity, mass-conserving superwind
flowing into a solid angle $\Omega_w$ at a velocity $\Delta v$ from a
minimum radius ($r_*$ - taken to be the radius of the starburst within which
the flow originates). This implies:

\begin{displaymath}
\dot{M} \sim 30 (r_*/kpc) (N_H/10^{21} cm^{-2}) (\Delta v/300 km/s)
 (\Omega_w/4\pi) M_{\odot}/yr
\end{displaymath}

\begin{displaymath}
\dot{E}  \sim 10^{42} (r_*/kpc) (N_H/10^{21} cm^{-2})
(\Delta v/300 km/s)^3  (\Omega_w/4\pi) erg/s
\end{displaymath}

Based on this simple model, H2000 estimate that the
implied outflow rates of cool atomic gas are comparable to the
star-formation rates (e.g. several tens of solar masses per year
in powerful starbursts). The flux of kinetic energy carried by
this material is substantial (of-order 10$^{-1}$ of the kinetic
energy supplied by the starburst).

In the best-studied outflows it has been possible to get rough estimates
of the metallicity of the cool atomic phase. In the high-redshift galaxy
MS 1512-cB58, Pettini et al (2000) obtain $\sim$1/4 solar, while
Heckman \& Leitherer (1997) find 20 to 50\% solar
in the outflow in the dwarf starburst NGC1705. These measures
are consistent with the theoretical expectation that the cool component
of the outflow is mostly ambient interstellar gas accelerated by the wind.

\begin{figure}
\plotone{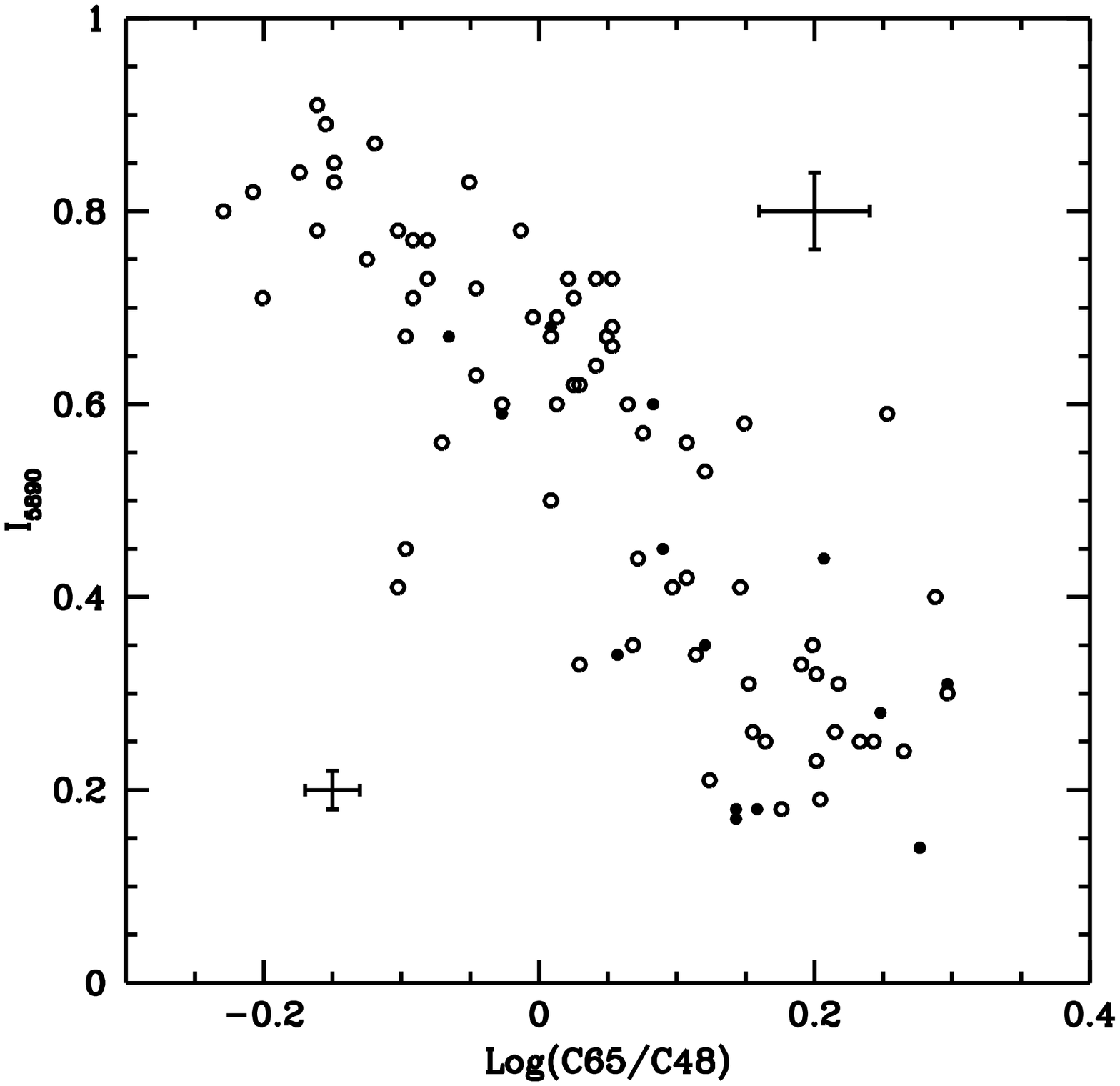}
\caption{\small Plot of the normalized residual intensity at the
center of the interstellar $NaD$ $\lambda$5890
absorption-line ($I_{5890}$) {\it vs.} the log of the color of the optical
continuum
(the ratio of $F_{\lambda}$ at rest wavelengths of 6560 and 4860
\AA). Points plotted as solid dots are the nuclei of infrared-bright starbursts
and the other points
are off-nuclear locations. See H2000 for details.
The deeper the
$NaD$ line (higher covering factor), the more-reddened the background starlight.
The correlation is obeyed by both the nuclear and off-nuclear regions.
An unreddened starburst population should have $log(C_{65}/C_{48})$
= -0.3. For a standard Galactic reddening curve, the implied
$A_V$ ranges up to roughly 4 magnitudes for the most-reddened sight-lines.
Typical uncertainties for the nuclear (extra-nuclear) data are indicated
by the error-bar in the lower-left (upper-right) of the plot.}
\label{fig2}
\end{figure}

A bigger surprise is that the $NaD$ survey of H2000
implies that a substantial amount of {\it dust} is being expelled
along with the atomic gas. Figure 2 show the strong correlation between
the depth of the blueshifted $NaD$ absorption-line (a measure of the
covering factor of the absorbing gas) and the line-of-sight reddening.
The implied dust outflow rates are substantial (of-order 10$^{-2}$
of the mass outflow rate, or typically 0.1 to 1 M$_{\odot}$ per year
in powerful starbursts).

\subsection{Summary of Outflow Rates}

In summary, the various techniques for estimating the outflow rates
in superwinds rely on simplifying assumptions (not all of which may be
warranted). On the other hand, it is gratifying that the different
techniques do seem to roughly agree: {\it the outflows carry
mass out of the starburst at a rate comparable to the star-formation
rate and kinetic/thermal energy out at a rate comparable to the rate
supplied by the starburst.}

Estimates of the rates at which metals and dust are carried out
are more uncertain. It appears that the metallicity
of a typical outflow is roughly solar, and that the dust-to-gas ratio
in the cool atomic component of the outflow is similar to the
Galactic value (e.g. dust $\sim$1\% by mass).

\section{The Fate of Superwinds}

The outflow rates in superwinds should not be taken directly
as the rates at which mass, metals, and energy {\it escape} from
galaxies and are transported into the intergalactic medium.
After all, the observable manifestations of the outflow are produced by
material still relatively deep within the gravitational potential
of the galaxy's dark matter halo.

One way of assessing the likely fate of the superwind material
is to compare the observed or estimated outflow velocity to
the estimated escape velocity from the galaxy (see Martin 1999;
H2000).
For an isothermal gravitational potential that extends to a maximum radius
$r_{max}$, and has a circular rotation velocity $v_{rot}$, the escape velocity
at a
radius $r$ is given by:

\begin{displaymath}
v_{esc} = [2v_{rot}(1 + ln(r_{max}/r))]^{1/2}
\end{displaymath}

In the case of the interstellar absorption-lines, H2000
argued that the observed profiles were produced by material
ablated off of ambient clouds at the systemic velocity and accelerated
by the wind up to a terminal
velocity represented by the most-blueshifted part of the profile.
In the case of the X-ray data, we do not measure a Doppler shift
directly, but we can define a characteristic outflow speed $v_X$
corresponding to the observed temperature $T_X$, assuming an adiabatic
wind with a mean mass per particle $\mu$ (see Chevalier \& Clegg 1985):

\begin{displaymath}
v_X \sim (5kT_X/\mu)^{1/2}
\end{displaymath}

This is a conservative approach as it ignores the kinetic energy the
X-ray-emitting gas already has (probably a factor
typically 2 to 3 times
its thermal energy - Strickland \& Stevens 2000).

The results of comparing the outflow and escape velocities are shown
in Figure 3 (see H2000), which suggests that the outflows can readily
escape from dwarf galaxies, but possibly not from the more massive
systems.

\begin{figure}
\plotone{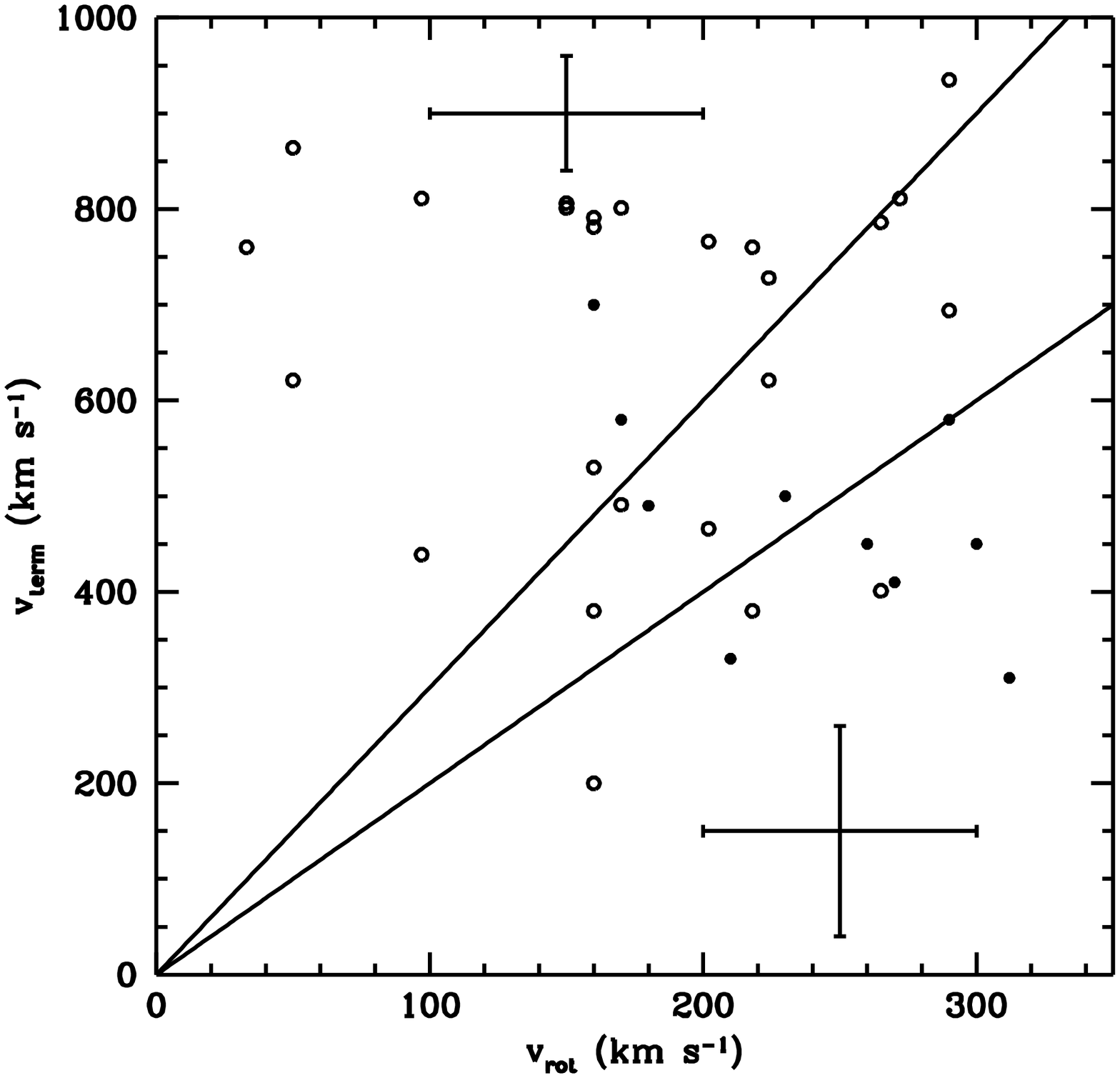}
\caption{\small Plot of the galaxy rotation speed
{\it vs.} the inferred terminal velocity of the outflows (see H2000). The
solid points are based on the $NaD$ absorption-line profiles.
The hollow points are estimated from
the observed temperature of the hot X-ray-emitting gas (see text).
Note that the two data sets are consistent with
each other,
imply that the outflow speed is independent of the host galaxy
potential well depth, and thus suggest that outflows will
preferentially escape from the least massive galaxies. The two diagonal
lines indicate the galaxy escape velocity under the assumption
that $v_{esc}$ = 2 $v_{rot}$ and $v_{esc}$ = 3 $v_{rot}$ respectively.
Typical uncertainties in the
X-ray ($NaD$) estimates of $v_{term}$ are shown by the error-bar on the
bottom right (upper center).}
\label{fig3}
\end{figure}

How far out from the starburst can the effects of superwinds be observed?
In general, such tenuous material will be better traced via absorption-lines
against background QSOs than by its emission (since the emission-measure
will drop much more rapidly with radius than will the column density). To date,
the only such experiment that has been conducted is by Norman et al (1996)
who examined two sight-lines through the halo of the merger/starburst
system NGC520 using HST to observe the MgII$\lambda$2800 doublet. Absorption
was definitely detected towards a QSO with an impact parameter of
35 $h_{70}^{-1}$ kpc and possibly towards a second QSO with 
an impact parameter of 75 $h_{70}^{-1}$ kpc. Since NGC520 is immersed
in tidal debris (as mapped in the HI 21cm line), it is unclear
whether the MgII absorption is due to tidally-liberated or wind-ejected
gas. We can expect the situation to improve in the next few years,
as the $Galex$ mission and the $Sloan~Digital~Sky~Survey$ provide us with $10^5$
new QSOs and starburst galaxies, and the $Cosmic~Origins~Spectrograph$
significantly improves the UV spectroscopic capabilities of HST.

While a wind's X-ray surface brightness drops rapidly with radius due
to expansion and adiabatic cooling, its presence at large radii can
be inferred if it collides with an obstacle. In the case of M82,
Lehnert, Heckman, \& Weaver (1999) show that a ridge of diffuse X-ray 
and H$\alpha$
emission at a projected distance of 12 kpc from the starburst is
most likely due to a wind/cloud collision in the galaxy halo. An
even more spectacular example (Irwin et al 1987) is the peculiar tail of
HI associated
with the galaxy NGC3073 which points directly away from
the nucleus of its companion: the prototypical superwind galaxy NGC3079
(60 $h_{70}^{-1}$ kpc away form NGC3073 in projection). Irwin et al (1987)
proposed that the HI tail is being swept out of NGC3073 by the 
ram pressure of NGC3079's superwind.
 
\section{Implications of Superwinds}

As discussed above, we now know that superwinds are ubiquitous in
actively-star-forming galaxies in both the local universe, and at
high-redshift. The outflows detected in the high-z Lyman Break galaxies
are particularly significant, since these objects may plausibly
represent the sites where the majority of the stars and metals
have been produced over the history of the universe (e.g. Adelberger
\& Steidel 2000). Even if the sub-mm $SCUBA$ sources turn out
to be a distinct population at high-z, their apparent similarity
to local ``ultraluminous galaxies'' suggests that they too
will drive powerful outflows (cf. Heckman et al 1996).
With this in mind, let me briefly describe the implications of superwinds
for the evolution of galaxies and the inter-galactic medium.

Figure 3 shows that the estimated outflow speeds in superwinds are $v_{wind}$
$\sim$ 300 to 800 km s$^{-1}$, and are independent of the rotation speed
of the ``host galaxy'' over the range $v_{rot}$ = 30 to 300 km s$^{-1}$.
This strongly suggests that the outflows selectively escape the potential
wells of the less massive galaxies. H2000 considered the simple
model based on Lynden-Bell (1992) in which the fraction
of starburst-produced metals
that are retained by a galaxy experiencing an outflow is proportional to the
galaxy potential-well
depth for galaxies with 
escape velocities $v_{esc} < v_{wind}$, and asymtotes to
full retention for the most massive galaxies ($v_{esc} > v_{wind}$).
For $v_{wind}$ in the above range, such a simple prescription
can reproduce the observed mass-metallicity relation for elliptical
galaxies.

The selective loss of gas-phase baryons from low-mass galaxies
via supernova-driven winds is an important ingredient in
semi-analytic models of galaxy formation (e.g. Somerville \& Primack 1999).
It is usually
invoked to enable the models to reproduce the observed faint-end
slope of the galaxy luminosity function by selectively suppressing
star-formation in low-mass dark-matter halos. A different
approach is taken by
Scannapieco, Ferrara, \& Broadhurst (2000), who have argued that
starburst-driven outflows
can suppress the formation
of dwarf galaxies by ram-pressure-stripping the
gaseous baryons from out of the dark-matter halos of low-mass {\it companion}
galaxies.
The NGC3073/3079
interaction (Irwin et al 1987) may represent a local example.

H2000 also show that the simple Lynden-Bell (1992) model
for a mass-dependent wind-driven loss of metals can
(if applied to the population of elliptical galaxies and bulges
in a cluster) deposit the required amount
of observed metals in the intra-cluster medium. If the ratio of ejected
metals to stellar
spheroid mass is the same globally as in clusters of galaxies,
we predicted that the present-day
mass-weighted metallicity of a general intergalactic medium
with $\Omega_{igm}$ = 0.015 will be $\sim$ 1/6 solar (see also
Renzini 1997).

There is now a consensus that the inter-galactic medium has been
significantly heated by a non-gravitational source (probably at rather
early epochs).
This ``pre-heating'' is required to explain the steep luminosity-temperature
relation for the X-ray gas in galaxy clusters (e.g. Tozzi \& Norman 2000),
the existence of an entropy excess in the gas in the central regions of small
clusters (Ponman, Cannon, \& Navarro 1999), and the relative faintness
of any contribution by the inter-galactic medium to the cosmic X-ray
background (e.g. Pen 1999). The required amount of heating depends
somewhat on the details and history of the heating process, but
amounts to of-order 1 keV per baryon. It is interesting that this
is just the amount produced by superwinds (assuming 100\% 
heating efficiency!).
A standard Salpeter IMF extending from 0.1 to 100 $M_{\odot}$
produces about 10$^{51}$ ergs of kinetic energy from supernovae
per 80 $M_{\odot}$ of low-mass stars ($\leq$ 1 $M_{\odot}$). The present
ratio of baryons in the intra-cluster medium to baryons in low-mass
stars is $\sim$ 6 in clusters, so the amount of kinetic energy 
available in principle
to heat the intra-cluster medium is then 10$^{51}$ ergs per
480 $M_{\odot}$, or $\sim$ 1 keV per baryon.

H2000 have summarized the evidence that starbursts are ejecting
significant quantities
of dust. {\it If}
this dust can survive a trip into the intergalactic medium and remain
intact for a Hubble time,
they estimated that the upper bound on the global amount of intergalactic
dust is $\Omega_{dust}$ $\sim$ 10$^{-4}$. While this is clearly an
upper limit, it is a cosmologically interesting one:
Aguirre (1999) argues that dust this abundant
could in principle obviate the need for a positive cosmological
constant, based on the Type Ia supernova Hubble diagram.

{\it It seems clear that superwinds play a fundamental role in the
evolution of galaxies and the inter-galactic medium.}

% For examples on including figures, see the file vla2000_sample.ps
% at http://www.nrao.edu/vla2000/proceedings/. 
% For examples of figures, equations or tables, please see the file
% vla2000_man.ps at the same site. Also available as
% newpaspman.ps at http://www.aspsky.org/pubs/authors.html

% comment this out if you want to include acknowledgements

\acknowledgements

I would like to thank my principal collaborators on the work described in this
contribution: L. Armus, M. Dahlem, R. Gonzalez-Delgado, M. Lehnert,
C. Leitherer, A. Marlowe, C. Martin, G. Meurer,
C. Norman, K. Sembach, D. Strickland, and K. Weaver.
This work has been supported
in part by grants from the NASA LTSA program and the $HST$, $ROSAT$, $ASCA$, and
$Chandra$ GO programs.

\end{document}